\documentclass[10pt,twoside]{article}
\usepackage{epsf,colordvi,color,amsbsy}
\usepackage[dvips]{graphicx}
\begin{document}

\title{\bfseries
Wavelet Cross-Correlation Analysis\
 of Turbulent Mixing\
 from Large-Eddy-Simulations}
%
\author{ S.~Sello and J.~Bellazzini}
\date{}
\maketitle
\vspace*{-9 mm}
{\small
\begin{center}
Enel Research\\
 A.~Pisano 120, Pisa 56122, ITALY
\end{center}
}
\noindent
Contact e-mail: sello@pte.enel.it
\thispagestyle{empty}
\section{Introduction}
The complex interactions existing between turbulence and mixing in a
bluff-body stabilised flame configuration is investigated by means
of a wavelet cross-correlation analysis on Large Eddy Simulations.
The combined approach allows to better point out typical features of unsteady
turbulent flows with mixing through the characterisation of the
processes involved both in time and scales.
The wavelet cross-correlation analysis of the time signals of
velocity and mixture fraction fluctuations can be an an effective tool
to study the processes involved in turbulent mixing flows which are of great
interest in combustion problems.

\section{Generalities on wavelet cross-correlation}
The continuous wavelet transform of a function $f(t)$ is defined as the
convolution between $f$ and a dilated function $\psi$ called wavelet mother:
\begin{equation}
W_{f}(a,\tau) = \frac{1}{\sqrt{a}} \int_{-\infty}^{+\infty}
f(t)\psi^* (\frac {t-\tau}{a})dt,
\end{equation}
where $a$ is  the dilation parameter, which plays the same role as the
frequency in Fourier analysis, and $\tau$ indicates the translation parameter
corresponding to the position of the wavelet in the physical space.
 In the present study we use the complex Morlet wavelet
($\psi(t) =  e^{i\omega_{0} t} e^{-t^{2}/2}$)  as wavelet mother.\\
Let $W_{f}(a,\tau)$ and $W_{g}(a,\tau)$ be the continuous wavelet transforms
of $f(t)$ and $g(t)$. We define the \emph{wavelet cross-scalogram} as
\begin{equation}
W_{fg}(a,\tau) = W_{f}^{*}(a,\tau)W_{g}(a,\tau),
\end{equation}
where the symbol $*$ indicates the complex conjugate. When the wavelet mother
is complex,
the wavelet cross-scalogram $W_{fg}(a,\tau)$ is also complex and can be
written in terms of its real and imaginary parts:
\begin{equation}
W_{fg}(a,\tau) = Co W_{fg}(a,\tau) -iQuad W_{fg}(a,\tau).
\end{equation}
It can be shown that the following equation holds if $f(t), g(t) \in
\mathcal{L}^{2}(\Re)$
\begin{equation}
\int_{-\infty}^{+\infty} f(t)g(t)dt  =  1/c_{\psi} \int_{0}^{+\infty}\int_
{-\infty}^{+\infty} Co W_{fg}(a,\tau)d\tau da,
\end{equation}
where $1/c_{\psi}$ is a constant depending on the choice of the wavelet mother.

\section{ Cross wavelet coherence functions}
The highly redundant information from a multiscale wavelet analysis of time
series must be reduced by means of suitable selective procedures and
quantities, in order to extract the main features correlated to an
essentially intermittent dynamics.
In this study, we analysed and compared the properties of two
complementary wavelet local correlation coefficents which are able to well
evidence peculiar and anomalous local events associated to the vortex dynamics.
More precisely, given two signals $f(t)$ and $g(t)$, we refer to the so-called
\emph{Wavelet Local Correlation Coefficent} (Buresti {\em et. al}~\cite{buresti}),
defined as:
\begin{equation}
WLCC(a,\tau) = \frac{Co W_{fg} (a,\tau)}{\mid W_{f}(a,\tau)\mid \mid
W_{g}(a,\tau)\mid}.
\end{equation}
This quantity is essentially a measure of the phase coherence of the signals.
Here we introduce the \emph{Cross Wavelet Coherence Function} (CWCF) defined
as:
\begin{equation}
CWCF(a,\tau)=\frac{2\mid W_{fg}(a,\tau)\mid^2}{\mid W_{f}(a,\tau)\mid^4 +\mid
W_{g}(a,\tau)\mid^4},
\end{equation}
which is essentially a measure of the intensity coherence of the signals.
Using the polar coordinates 
we can write the wavelet transforms of
$W_{f}(a,\tau)$, $W_{g}(a,\tau)$ and $W_{fg}(a,\tau)$ as:
\begin{equation}
W_{f}(a,\tau) = \rho_{f}e^{\imath \theta_{f}} \hspace{0.5cm}
W_{g}(a,\tau) = \rho_{g}e^{\imath \theta_{g}}
\end{equation}
\begin{equation}
W_{fg}(a,\tau) = \rho_{f}\rho_{g}e^{\imath (\theta_{g}-\theta_{f})},
\end{equation}
and the Cross Wavelet Coherence Function can be written also as:
\begin{equation}
CWCF(a,\tau) = \frac{2\rho_{f}^{2}\rho_{g}^{2}}{\rho_{f}^{4}+\rho_{g}^{4}}.
\end{equation}
It is easy to observe the two basic properties of the function (6):
\begin{equation}
CWCF(a,\tau) = 0  \Longrightarrow \rho_{f} = 0\qquad \textrm{or } \rho_{g} =  0
\end{equation}
\begin{equation}
0\leq CWCF \leq 1\qquad \forall ~a, \tau.
\end{equation}

\section{Numerical simulation}
We considered a laboratory-scale axisymmetric flame of methane-air in a non
confined bluff-body configuration. More precisely, the burner consists of a
5.4 mm diameter methane jet located in the center of a 50 mm diameter
cylinder. Air is supplied through a 100 mm outer diameter coaxial jet around
the 50 mm diameter bluff-body.
The Reynolds number of the central jet is 7000 (methane velocity
 =21 m/s) whereas the Reynolds number of the coaxial jet is 80000
(air velocity =25 m/s). This is a challenging test case for all the turbulence
models, as well documented in the ERCOFTAC report (Chatou, 1994)~\cite{ercoftac}. Moreover, due
to the highly intermittent, unsteady dynamics involved and the high turbulence level, especially for the reactive case, the Large Eddy Simulation (LES) appears as the most adequate numerical approach (Sello {\em et. al}~\cite{sello}).

\section{Results and discussion}
In this analysis we are mainly interested to relations existing between evolution of turbulence and mixing, for the reactive case. Previous DNS simulations on coaxial jets at different Reynolds numbers, show the ability of the wavelet cross-correlation analysis to better investigate the relations between mixing process and the dynamics of vorticity (Salvetti {\em et. al}~\cite{salvetti}). Thus, the signals analysed here are velocity fluctuations (for Reynolds stress contributions) and mixture fraction fluctuations (for mixing evolution) from LES.
As an example, Figure~1 shows the wavelet co-spectrum maps for a significant time interval in the pseudo-stationary regime of motion.
The main contributions to the Reynolds stress are evidenced by high intensity correlations (red) and anti-correlations (blue) regions, which evolve intermittently.
The dominant frequencies involved are located around 130 Hz.
For the mechanisms responsable of the evolution of mixing, we note that the same regions of high Reynolds stress correspond to high correlation, or cooperation, between velocity and mixture fraction fluctuations, suggesting that, at the selected location, the same events of stretching and tilting of the vorticity layer, drive both Reynolds stress and mixing evolutions.
Note that the large high value region located at low frequencies in the right map
is statistically not significant if we assume a proper red noise background 
spectrum.
To better investigate the role of the high correlation regions, we performed a cross section in the wavelet map at the frequency 160 Hz. Figure~2 (left) shows the time behaviour of the coherence functions WLCC, eq.(5), and CWCF, eq.(6). Here the phase and intensity coherence of signals are almost equivalent, but we can clearly point out an important anomalous event occurred at around t=0.19 s, corresponding to a loss of both intensity and phase coherence, followed by a change of the correlation sign. 
The link between this event and the dynamics of vorticity is evidenced by Figure~2 (right), which displays the wavelet map of the related vorticity signal.
The higher frequency significant regions ($\approx$ 730 Hz) result strongly intermittent, with a bifurcation to lower and higher values than average, followed by a drop of activity, in phase with the anomalous event.
\begin{figure}[!h]
\begin{minipage}[c]{.39 \linewidth}
\end{minipage}
\hfill
\begin{minipage}[c]{.39 \linewidth}
\begin{center}
\end{center}
\end{minipage}
\caption{Cross-Wavelet co-spectrum maps for axial and radial velocity fluctuations
(left) and for axial velocity and mixture fraction fluctuations
(right) at a given spatial point near the edge of the central jet.}
\end{figure}

\begin{figure}[!h]
\begin{minipage}[c]{.39 \linewidth}
\end{minipage}
\hfill
\begin{minipage}[c]{.39 \linewidth}
\begin{center}
\end{center}
\end{minipage}
\caption{Coherence functions for axial velocity and mixture fraction fluctuations
(left) and wavelet map of vorticity time series (right).} 
\end{figure}
These few examples support the usefulness of the cross-wavelet analysis approach to better investigate turbulent mixing processes in real systems.

\end{document}